\documentclass[%
reprint,
aip,
jmp,%
amsmath,amssymb,
]{revtex4-1}

\usepackage{graphicx}
\usepackage{dcolumn}
\usepackage{bm}
\usepackage{natbib} 

\begin{document}

\preprint{AIP/123-QED}

\title{Injection-locking of a fiber-pigtailed laser to an external cavity diode laser via fiber optic circulator}
\author{Toshihiko Shimasaki}    
\altaffiliation[Present address: ]{Department of Physics, University of California Santa Barbara, California 93106 USA}
\altaffiliation[]{toshihiko.shimasaki@gmail.com} 
\author{Eustace R. Edwards}
\author{David DeMille}
\affiliation{
Department of Physics, Yale University, New Haven, Connecticut 06520, USA\
}



%
\date{\today}
\begin{abstract}
We report a simple tunable master-slave laser injection-lock scheme for atomic physics experiments.
Seed light from an external cavity diode laser is injected into a high-power fiber-pigtailed diode laser via a fiber optic circulator. High-power outputs (up to $\sim$600 mW) at the injected frequency have been obtained in a single-mode fiber with tuning over a wide wavelength range ($\sim$15 nm). The scheme is simpler and more cost-effective than the traditional scheme of free-space injection-locking.  
\end{abstract} 

\pacs{Valid PACS appear here}
\maketitle


Tunable single frequency external cavity diode lasers (ECDLs) are widely used in atomic physics experiments.\cite{Wieman1991} Typically, $\lesssim$100 mW of output power can be obtained from an ECDL. When higher powers are required, master-slave setups based on injection-locking are frequently used to reach higher power output while maintaining the tuning and linewidth characteristics of the ECDL. In these systems, typically seed light from a master laser is coupled via free space into a slave laser, with the combination of a Faraday rotator and a polarizing beamsplitter acting to separate the amplified beam from the injected beam.\cite{stover1966locking} 
Such free-space injection-locking, however, is often susceptible to unlocking of the slave laser from the injected frequency. The success of the injection locking depends on the mode matching between the seed beam and the slave diode, on the in-coupled seed power, and on the matching between the seed frequency and the internal cavity resonance of the seed laser. 
Changes in the operating conditions over time can result in unreliable locking of the slave diode. An interesting and practical approach to circumvent such unlocking has been reported recently.\cite{Gupta2016} 
Alternatively, tapered amplifiers (TAs) can also be used to boost the available power. The power available from a TA depends on the wavelength, and is quite high at a few commonly-used wavelengths such as 850 nm. 
However, the range of wavelengths accessible with TAs is limited, their cost is substantial, and they often require free-space coupling as well. In addition, the beam quality obtained from a TA is often poor, leading to a low coupling efficiency of the TA output beam to a single-mode optical fiber. 

\begin{figure}[htp]
\includegraphics[scale=1]{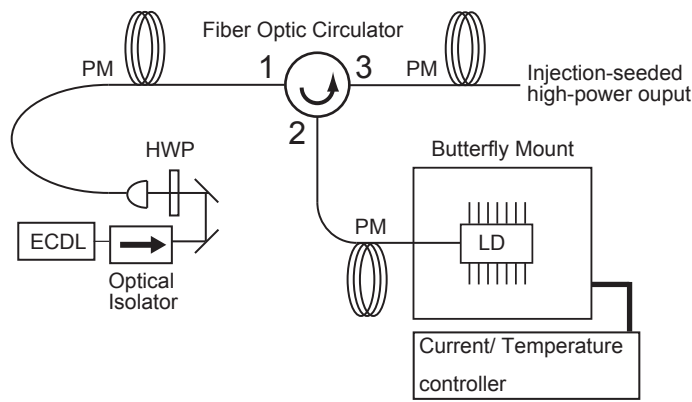}
\caption{\label{fig:Scheme}Schematic of the experimental setup. ECDL, external cavity diode laser; HWP, half wave plate; LD, fiber-pigtailed high power laser diode; PM, polarization-maintaining single-mode fiber.}
\end{figure}

In this paper, we demonstrate a simple scheme to obtain moderate powers, up to 600 mW at 980 nm, of tunable single-frequency laser light directly in an optical fiber. 
Our scheme uses a fiber-pigtailed laser diode as a slave laser, injection-locked by a seed from an ECDL via a fiber optic circulator. 
Injection-locking via a fiber optic circulator has been demonstrated at 1550 nm previously.\cite{Chrostowski2006} However, this approach has not, to our knowledge, been used before in visible or near infrared wavelengths. 
Recently, it has become possible to obtain fiber optic circulators for shorter wavelengths, with specifications comparable to those for free-space isolators. By using a fiber optic circulator, we eliminate most of the free-space components involved in a typical injection locking setup, and hence reduce the alignment sensitivity. The only free-space components are those used to frequency-stabilize an ECDL and to couple the output of the ECDL to an optical fiber. This greatly increases the reliability of the system. The simplicity of the setup also lowers the cost and required space for the setup. 
A schematic of our setup is shown in Fig. \ref{fig:Scheme}. 

Injection seed light from an ECDL is coupled to a polarization-maintaining (PM) fiber and delivered to Port 1 of a fiber optic circulator (Opneti, HPMCIR-980-F-900-0.8-FA-1W) through a fiber-to-fiber sleeve (Thorlabs, ADAFCPM2). (We use fiber-to-fiber sleeves for the sake of easy troubleshooting. The loss at the connector is typically 0.5 dB and could be reduced further with index matching gel.) 
The ECDL has to be coupled to the operation axis (slow axis in our case) of the circulator. The insertion loss of the circulator for the laser output is specified to be 2.6 dB. 
The seed beam is taken out from Port 2 of the circulator and injected into a fiber-pigtailed high power laser diode specified to operate at 976 nm (Gooch \& Housego, AC1401-0600-0976-00) that is used as the slave laser. 
This laser can be purchased with an optional fiber Bragg grating (FBG). However, because the frequency-dependent transmission of the FBG can possibly interfere with injection locking, we use a version of the laser without a FBG as our slave laser. 
In the free-running condition, this laser can deliver up to 800 mW. The output of the slave laser exits from Port 3 of the circulator, where up to 600 mW is available as a single-frequency, tunable output from the single-mode PM fiber. 

\begin{figure}
\includegraphics[scale=1]{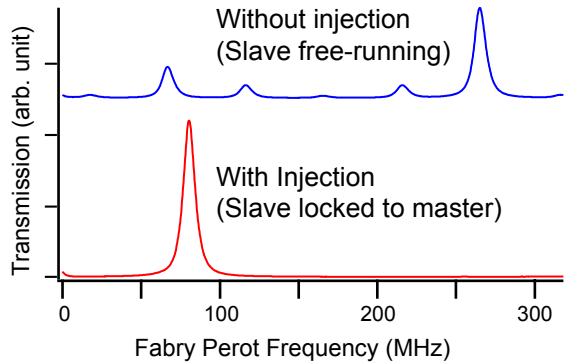}
\caption{\label{fig:FP} Transmission signal through a scanning Fabry Perot cavity. Here, the frequency resolution is limited by the FSR and finesse of the cavity.}
\end{figure}

Fig. \ref{fig:FP} shows the output spectrum of the slave with and without injection. 
The spectrum for operation without injection shows multiple peaks due to the multi-longitudinal mode output of the slave. 
In contrast, the spectrum with injection shows a single peak corresponding to single-mode operation at the injected frequency. If the injected power is not sufficient, obtaining such a single-mode output while tuning the frequency of the seed laser requires occasional minor re-tuning of the slave current to achieve matching of the slave cavity frequency to the injected frequency; this behavior is common in any injection locking system.\cite{Gupta2016} By operating with fairly high injection power, up to $\sim$50 mW, we were able to operate the slave laser in the locked condition over the entire range 970 nm - 985 nm. We believe that this is a result of good mode-matching between the seed beam and the slave diode and low reflectivity of the laser diode at the chip facet. Similar behavior has been reported in a free-space injection-locking of an anti-reflection coated diode, where an injection power up to 20\% of the slave laser power was used.\cite{ShimadaSrLaser} However, AR-coated diodes typically deliver lower output power than the same diode without AR coating.

To systematically investigate the locking stability, we examined the injection-locking bandwidth as a function of the slave output power and the injection power. Theoretically, the relation between the locking bandwidth $\Delta f$ (i.e., the range over which the seed laser can be tuned while maintaining the injection lock) and the ratio of the slave output power to the injected power is given by \cite{siegman1986lasers,Mogensen1985}
\begin{equation}   
\Delta f = \frac{1}{2\pi}\eta \frac{c}{2 n L} \sqrt{\frac{P_i}{P_o}} \sqrt{1+\alpha^2}, \label{eq:LockWidth}
\end{equation}
where $\eta$ is the injection coupling efficiency ($0\le \eta \le 1$), $c$ is the speed of light, $n$ is the refractive index inside the diode cavity, $L$ is the effective cavity length, and $P_i$ and $P_o$ are the injected seed power and the slave output power measured outside the laser diode, and $\alpha$ is a property of the semiconductor gain medium known as the linewidth enhancement factor.\cite{Henry1982} Here, $\eta$ includes factors such as imperfect coupling of the injected light to the slave and the transmission at the facet of the diode chip; for our system, we estimate $\eta \approx 1$. 
We do not know the value of $\alpha$ for our system. 
However, large values (in the range 3-10 or even up to $\sim$50) have been reported in other
semiconductor laser systems\cite{InGaAs_alpha1996,InGaAsQW2001}. Hence, it is plausible that, in our system, the locking bandwidth $\Delta f$ given by Eq. \ref{eq:LockWidth} could be comparable to or larger than the laser free spectral range (FSR), $\delta f_L = c/2nL$. 
In such situations, multiple longitudinal modes become relevant when the injection frequency is scanned over the range of the locking width and the the model used to derive Eq. \ref{eq:LockWidth} is no longer complete. 
That is, the slave lasing mode transitions from one longitudinal cavity mode to the next when the injection frequency is swept over more then the FSR of the slave diode; the slave lases in the longitudinal mode that is closest to the injection frequency and this mode is injection-locked.
This should make the effective locking width many times larger than that given in Eq. \ref{eq:LockWidth}, only limited by other parameters such as the gain profile of the slave laser. 

\begin{figure}
\includegraphics[scale=0.48]{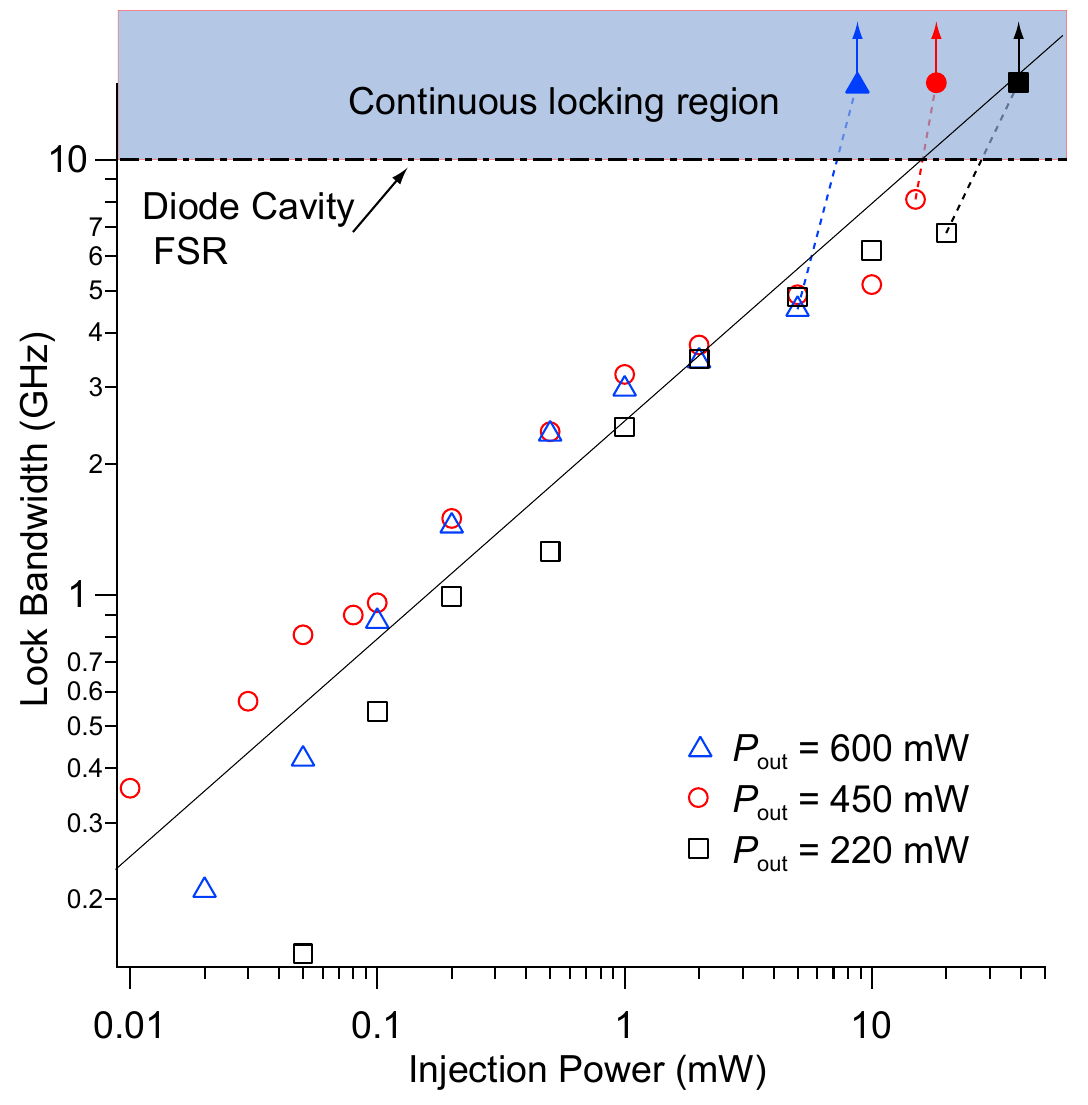}
\caption{\label{fig:LockWidth} Injection locking bandwidth as a function of the injection power (measured at Port 1 of the circulator). Three data sets with different output power (measured at Port 3 of the circulator) are shown. A solid line indicating the slope predicted by Eq. \ref{eq:LockWidth} is inserted as a guide for the eye. For each operating current of the slave laser, when the injection power is above a certain threshold, the locking bandwidth exceeds the FSR of the slave diode ($\sim$10 GHz, indicated by a dashed dotted line) and the slave laser follows the master laser seamlessly.
In this regime of operation, the slave laser stayed locked to the seed over the entire $\sim$20 GHz continuous scan range available with the piezo drive of our ECDL, and even when the master laser frequency was shifted by a significant amount ($\sim$300 GHz or more) by manually changing the grating angle of the ECDL.
The locking bandwidth in this continuous locking regime is denoted by the shaded area to indicate a locking bandwidth of $\gg$10 GHz. Data points in this region are shown with filled markers. 
}
\end{figure}

Fig. \ref{fig:LockWidth} shows the measured locking bandwidth as a function of the injection power, for three operating powers of the slave laser. For each slave current and injected optical power, we measured the locking bandwidth by sweeping the master laser frequency. 
At higher injection powers, the locking bandwidth exceeds the FSR of the slave diode (determined in auxiliary measurements to be $\delta f_L\sim$10 GHz), and we confirm that the master laser can be tuned over a wide frequency range ($\gg 10$ GHz) without the slave laser coming unlocked, as expected from the argument above. 
From a practical point of view, this broad range of unlock-free operation is very useful. 



We also measured the beat spectrum between the master laser and the slave laser (after offsetting the slave with an acousto-optic modulator), in order to check if there is any linewidth broadening due to the amplification in the slave laser. The linewidth of the master-slave beat spectrum was less than 100 kHz, which is roughly the free-running short-time linewidth of the master laser. This indicates that our injection-locking scheme does not significantly broaden the laser linewidth. 
We were able to use this injection-locked laser to perform spectroscopy on molecular transitions in RbCs,\cite{Shimasaki2018RbCs} with typical linewidths of $\sim$10 MHz. 

Currently, similar high-power fiber-pigtailed diodes at many wavelengths, including 808 nm and 1064 nm, can be obtained commercially. We hope that similar devices will be available for a growing number of wavelengths in the future. It is also possible to obtain equivalent fiber optic circulators for other wavelengths. Combining this injection-lock scheme with a fiber-pigtailed DFB laser diode as the seed laser can eliminate any free-space component in the system. Such all-fiber systems would be very useful for space applications.

In summary, we have demonstrated a simple, cost-effective scheme for injection-locking of a high-power fiber-pigtailed laser diode to a tunable seed laser, via a fiber optic circulator. With about 10 mW of seed laser power from an ECDL, single-frequency, tunable output power of up to 600 mW has been obtained out of a single-mode fiber, with an operating wavelength range of more than 15 nm and very robust frequency-locking characteristics.


This work was supported by ARO MURI. We thank M.A. Bellos for providing useful comments, and we thank J.T. Kim and Y. Zhu for building the ECDL used in the experiment. TS acknowledges support from Yale University.
\bibliography{FiberInjectionLockPaper}

\begin{thebibliography}{11}%
\makeatletter
\providecommand \@ifxundefined [1]{%
 \@ifx{#1\undefined}
}%
\providecommand \@ifnum [1]{%
 \ifnum #1\expandafter \@firstoftwo
 \else \expandafter \@secondoftwo
 \fi
}%
\providecommand \@ifx [1]{%
 \ifx #1\expandafter \@firstoftwo
 \else \expandafter \@secondoftwo
 \fi
}%
\providecommand \natexlab [1]{#1}%
\providecommand \enquote  [1]{``#1''}%
\providecommand \bibnamefont  [1]{#1}%
\providecommand \bibfnamefont [1]{#1}%
\providecommand \citenamefont [1]{#1}%
\providecommand \href@noop [0]{\@secondoftwo}%
\providecommand \href [0]{\begingroup \@sanitize@url \@href}%
\providecommand \@href[1]{\@@startlink{#1}\@@href}%
\providecommand \@@href[1]{\endgroup#1\@@endlink}%
\providecommand \@sanitize@url [0]{\catcode `\\12\catcode `\$12\catcode
  `\&12\catcode `\#12\catcode `\^12\catcode `\_12\catcode `\%12\relax}%
\providecommand \@@startlink[1]{}%
\providecommand \@@endlink[0]{}%
\providecommand \url  [0]{\begingroup\@sanitize@url \@url }%
\providecommand \@url [1]{\endgroup\@href {#1}{\urlprefix }}%
\providecommand \urlprefix  [0]{URL }%
\providecommand \Eprint [0]{\href }%
\providecommand \doibase [0]{http://dx.doi.org/}%
\providecommand \selectlanguage [0]{\@gobble}%
\providecommand \bibinfo  [0]{\@secondoftwo}%
\providecommand \bibfield  [0]{\@secondoftwo}%
\providecommand \translation [1]{[#1]}%
\providecommand \BibitemOpen [0]{}%
\providecommand \bibitemStop [0]{}%
\providecommand \bibitemNoStop [0]{.\EOS\space}%
\providecommand \EOS [0]{\spacefactor3000\relax}%
\providecommand \BibitemShut  [1]{\csname bibitem#1\endcsname}%
\let\auto@bib@innerbib\@empty
\bibitem [{\citenamefont {Wieman}\ and\ \citenamefont
  {Hollberg}(1991)}]{Wieman1991}%
  \BibitemOpen
  \bibfield  {author} {\bibinfo {author} {\bibfnamefont {C.~E.}\ \bibnamefont
  {Wieman}}\ and\ \bibinfo {author} {\bibfnamefont {L.}~\bibnamefont
  {Hollberg}},\ }\bibfield  {title} {\enquote {\bibinfo {title} {{Using diode
  lasers for atomic physics}},}\ }\href {\doibase 10.1063/1.1142305} {\bibfield
   {journal} {\bibinfo  {journal} {Rev. Sci. Instrum.}\ }\textbf {\bibinfo
  {volume} {62}},\ \bibinfo {pages} {1--20} (\bibinfo {year}
  {1991})}\BibitemShut {NoStop}%
\bibitem [{\citenamefont {Stover}\ and\ \citenamefont
  {Steier}(1966)}]{stover1966locking}%
  \BibitemOpen
  \bibfield  {author} {\bibinfo {author} {\bibfnamefont {H.~L.}\ \bibnamefont
  {Stover}}\ and\ \bibinfo {author} {\bibfnamefont {W.~H.}\ \bibnamefont
  {Steier}},\ }\bibfield  {title} {\enquote {\bibinfo {title} {Locking of laser
  oscillators by light injection},}\ }\href {\doibase 10.1063/1.1754502}
  {\bibfield  {journal} {\bibinfo  {journal} {Applied Physics Letters}\
  }\textbf {\bibinfo {volume} {8}},\ \bibinfo {pages} {91--93} (\bibinfo {year}
  {1966})}\BibitemShut {NoStop}%
\bibitem [{\citenamefont {Saxberg}, \citenamefont {Plotkin-Swing},\ and\
  \citenamefont {Gupta}(2016)}]{Gupta2016}%
  \BibitemOpen
  \bibfield  {author} {\bibinfo {author} {\bibfnamefont {B.}~\bibnamefont
  {Saxberg}}, \bibinfo {author} {\bibfnamefont {B.}~\bibnamefont
  {Plotkin-Swing}}, \ and\ \bibinfo {author} {\bibfnamefont {S.}~\bibnamefont
  {Gupta}},\ }\bibfield  {title} {\enquote {\bibinfo {title} {{Active
  stabilization of a diode laser injection lock}},}\ }\href {\doibase
  10.1063/1.4953589} {\bibfield  {journal} {\bibinfo  {journal} {Rev. Sci.
  Instrum.}\ }\textbf {\bibinfo {volume} {87}},\ \bibinfo {pages} {063109}
  (\bibinfo {year} {2016})}\BibitemShut {NoStop}%
\bibitem [{\citenamefont {Chrostowski}\ \emph {et~al.}(2006)\citenamefont
  {Chrostowski}, \citenamefont {Zhao}, \citenamefont {Chang-Hasnain},
  \citenamefont {Shau}, \citenamefont {Ortsiefer},\ and\ \citenamefont
  {Amann}}]{Chrostowski2006}%
  \BibitemOpen
  \bibfield  {author} {\bibinfo {author} {\bibfnamefont {L.}~\bibnamefont
  {Chrostowski}}, \bibinfo {author} {\bibfnamefont {X.}~\bibnamefont {Zhao}},
  \bibinfo {author} {\bibfnamefont {C.~J.}\ \bibnamefont {Chang-Hasnain}},
  \bibinfo {author} {\bibfnamefont {R.}~\bibnamefont {Shau}}, \bibinfo {author}
  {\bibfnamefont {M.}~\bibnamefont {Ortsiefer}}, \ and\ \bibinfo {author}
  {\bibfnamefont {M.~C.}\ \bibnamefont {Amann}},\ }\bibfield  {title} {\enquote
  {\bibinfo {title} {{50-GHz optically injection-locked 1.55 $\mu$m VCSELs}},}\
  }\href {\doibase 10.1109/LPT.2005.862370} {\bibfield  {journal} {\bibinfo
  {journal} {IEEE Photonics Technology Letters}\ }\textbf {\bibinfo {volume}
  {18}},\ \bibinfo {pages} {367--369} (\bibinfo {year} {2006})}\BibitemShut
  {NoStop}%
\bibitem [{\citenamefont {Shimada}\ \emph {et~al.}(2013)\citenamefont
  {Shimada}, \citenamefont {Chida}, \citenamefont {Ohtsubo}, \citenamefont
  {Aoki}, \citenamefont {Takeuchi}, \citenamefont {Kuga},\ and\ \citenamefont
  {Torii}}]{ShimadaSrLaser}%
  \BibitemOpen
  \bibfield  {author} {\bibinfo {author} {\bibfnamefont {Y.}~\bibnamefont
  {Shimada}}, \bibinfo {author} {\bibfnamefont {Y.}~\bibnamefont {Chida}},
  \bibinfo {author} {\bibfnamefont {N.}~\bibnamefont {Ohtsubo}}, \bibinfo
  {author} {\bibfnamefont {T.}~\bibnamefont {Aoki}}, \bibinfo {author}
  {\bibfnamefont {M.}~\bibnamefont {Takeuchi}}, \bibinfo {author}
  {\bibfnamefont {T.}~\bibnamefont {Kuga}}, \ and\ \bibinfo {author}
  {\bibfnamefont {Y.}~\bibnamefont {Torii}},\ }\bibfield  {title} {\enquote
  {\bibinfo {title} {{A simplified 461-nm laser system using blue laser diodes
  and a hollow cathode lamp for laser cooling of Sr}},}\ }\href {\doibase
  10.1063/1.4808246} {\bibfield  {journal} {\bibinfo  {journal} {Rev. Sci.
  Instrum.}\ }\textbf {\bibinfo {volume} {84}},\ \bibinfo {pages} {063101}
  (\bibinfo {year} {2013})}\BibitemShut {NoStop}%
\bibitem [{\citenamefont {Siegman}(1986)}]{siegman1986lasers}%
  \BibitemOpen
  \bibfield  {author} {\bibinfo {author} {\bibfnamefont {A.~E.}\ \bibnamefont
  {Siegman}},\ }\href@noop {} {\emph {\bibinfo {title} {Lasers}}}\ (\bibinfo
  {publisher} {University Science Books},\ \bibinfo {year} {1986})\BibitemShut
  {NoStop}%
\bibitem [{\citenamefont {Mogensen}, \citenamefont {Olesen},\ and\
  \citenamefont {Jacobsen}(1985)}]{Mogensen1985}%
  \BibitemOpen
  \bibfield  {author} {\bibinfo {author} {\bibfnamefont {F.}~\bibnamefont
  {Mogensen}}, \bibinfo {author} {\bibfnamefont {H.}~\bibnamefont {Olesen}}, \
  and\ \bibinfo {author} {\bibfnamefont {G.}~\bibnamefont {Jacobsen}},\
  }\bibfield  {title} {\enquote {\bibinfo {title} {Locking conditions and
  stability properties for a semiconductor laser with external light
  injection},}\ }\href {\doibase 10.1109/JQE.1985.1072760} {\bibfield
  {journal} {\bibinfo  {journal} {IEEE Journal of Quantum Electronics}\
  }\textbf {\bibinfo {volume} {21}},\ \bibinfo {pages} {784--793} (\bibinfo
  {year} {1985})}\BibitemShut {NoStop}%
\bibitem [{\citenamefont {Henry}(1982)}]{Henry1982}%
  \BibitemOpen
  \bibfield  {author} {\bibinfo {author} {\bibfnamefont {C.}~\bibnamefont
  {Henry}},\ }\bibfield  {title} {\enquote {\bibinfo {title} {Theory of the
  linewidth of semiconductor lasers},}\ }\href {\doibase
  10.1109/JQE.1982.1071522} {\bibfield  {journal} {\bibinfo  {journal} {IEEE
  Journal of Quantum Electronics}\ }\textbf {\bibinfo {volume} {18}},\ \bibinfo
  {pages} {259--264} (\bibinfo {year} {1982})}\BibitemShut {NoStop}%
\bibitem [{\citenamefont {Bossert}\ and\ \citenamefont
  {Gallant}(1996)}]{InGaAs_alpha1996}%
  \BibitemOpen
  \bibfield  {author} {\bibinfo {author} {\bibfnamefont {D.~J.}\ \bibnamefont
  {Bossert}}\ and\ \bibinfo {author} {\bibfnamefont {D.}~\bibnamefont
  {Gallant}},\ }\bibfield  {title} {\enquote {\bibinfo {title} {{Gain,
  refractive index, and $\alpha$-parameter in InGaAs-GaAs SQW broad-area
  lasers}},}\ }\href {\doibase 10.1109/68.481104} {\bibfield  {journal}
  {\bibinfo  {journal} {IEEE Photonics Technology Letters}\ }\textbf {\bibinfo
  {volume} {8}},\ \bibinfo {pages} {322--324} (\bibinfo {year}
  {1996})}\BibitemShut {NoStop}%
\bibitem [{\citenamefont {Stohs}\ \emph {et~al.}(2001)\citenamefont {Stohs},
  \citenamefont {Bossert}, \citenamefont {Gallant},\ and\ \citenamefont
  {Brueck}}]{InGaAsQW2001}%
  \BibitemOpen
  \bibfield  {author} {\bibinfo {author} {\bibfnamefont {J.}~\bibnamefont
  {Stohs}}, \bibinfo {author} {\bibfnamefont {D.~J.}\ \bibnamefont {Bossert}},
  \bibinfo {author} {\bibfnamefont {D.~J.}\ \bibnamefont {Gallant}}, \ and\
  \bibinfo {author} {\bibfnamefont {S.~R.~J.}\ \bibnamefont {Brueck}},\
  }\bibfield  {title} {\enquote {\bibinfo {title} {{Gain, refractive index
  change, and linewidth enhancement factor in broad-area GaAs and InGaAs
  quantum-well lasers}},}\ }\href {\doibase 10.1109/3.958374} {\bibfield
  {journal} {\bibinfo  {journal} {IEEE Journal of Quantum Electronics}\
  }\textbf {\bibinfo {volume} {37}},\ \bibinfo {pages} {1449--1459} (\bibinfo
  {year} {2001})}\BibitemShut {NoStop}%
\bibitem [{\citenamefont {{Shimasaki}}\ \emph {et~al.}(2018)\citenamefont
  {{Shimasaki}}, \citenamefont {{Kim}}, \citenamefont {{Zhu}},\ and\
  \citenamefont {{DeMille}}}]{Shimasaki2018RbCs}%
  \BibitemOpen
  \bibfield  {author} {\bibinfo {author} {\bibfnamefont {T.}~\bibnamefont
  {{Shimasaki}}}, \bibinfo {author} {\bibfnamefont {J.-T.}\ \bibnamefont
  {{Kim}}}, \bibinfo {author} {\bibfnamefont {Y.}~\bibnamefont {{Zhu}}}, \ and\
  \bibinfo {author} {\bibfnamefont {D.}~\bibnamefont {{DeMille}}},\ }\bibfield
  {title} {\enquote {\bibinfo {title} {{Continuous Production of Rovibronic
  Ground State RbCs Molecules via Short-Range Photoassociation to the
  $b^3\Pi_{1}-c^3\Sigma^+_{1}-B^1\Pi_1$ States}},}\ }\href@noop {} {\bibfield
  {journal} {\bibinfo  {journal} {ArXiv e-prints}\ } (\bibinfo {year}
  {2018})},\ \Eprint {http://arxiv.org/abs/1802.01797} {arXiv:1802.01797
  [physics.atom-ph]} \BibitemShut {NoStop}%
\end{thebibliography}%
\end{document}